## REVIEW

**Open Access**

# Contemporary research trends in response robotics

Mehdi Dadvar[1*†]  and Soheil Habibian[2†]

**Abstract**

The multidisciplinary nature of response robotics has brought about a diversified research community with extended expertise. Motivated by the recent accelerated rate of publications in the field, this paper analyzes the research trends, statistics, and implications of the literature from bibliometric standpoints. The aim is to study the global progress of response robotics research and identify the contemporary trends. To that end, we investigated the collaboration mapping together with the citation network to formally recognize impactful and contributing authors, publications, sources, institutions, funding agencies, and countries. We found how natural and human-made disasters contributed to forming productive regional research communities, while there are communities that only view response robotics as an application of their research. Furthermore, through an extensive discussion on the bibliometric results, we elucidated the philosophy behind research priority shifts in response robotics and presented our deliberations on future research directions.

**Keywords:** Bibliometric analysis, Response robotics, Search and rescue robots, Collaboration mapping, Research trends

## Introduction

Response robotics (RR) leverages scientific and technological advancements from multiple engineering disciplines to enhance the efficacy and agility of rescue operations in natural and human-made disasters [1]. In the past few decades, researchers have explored a broad range of ground [2], aquatic [3], and aerial [4] solutions to address formidable challenges in harsh and disastrous environments. In spite of technological, scientific, and operational impediments, response robots have demonstrated notable effectiveness in the assigned missions in various catastrophes including but not limited to earthquakes [5], tsunamis [6], floods [7], nuclear incidents [8], fires [9], and terror attacks [10], as Fig. 1 demonstrates some remarkable examples of deployed response robots. Motivated by the interdisciplinary nature of RR overlapping with several other academic fields such as computer science, mechatronics, mechanical and electrical engineering, and cognitive science, this work aims to identify the global research and development trends in RR from a bibliometric perspective.

Although response robots have been developed with a wide variety of capabilities and technical specifications, they all have the principal capabilities in common: maneuvering, mobility, exploration, reconnaissance, and dexterity [11]. Depending on the nature of a mission, a response robot may have different combinations of the principal capabilities. As a result, response robots can be categorized according to their (1) size: mini-sized, man-packable, man-portable, and maxi-sized; (2) operational environment: ground, aquatic, and aerial; (3) locomotion mechanism: tracked, wheeled, legged, crawling, and bladed; (4) level of autonomy: tele-operative, semi-autonomous, and autonomous; and (5) collaborative skill: single-agent, and multi-agent. Each of these categories benefits the response operation in a specific set of tasks.

*Correspondence: mdadvar@lamar.edu
†Mehdi Dadvar and Soheil Habibian contributed equally to this work
[1] Was with Department of Computer Science, Lamar University, Beaumont, USA
Full list of author information is available at the end of the article

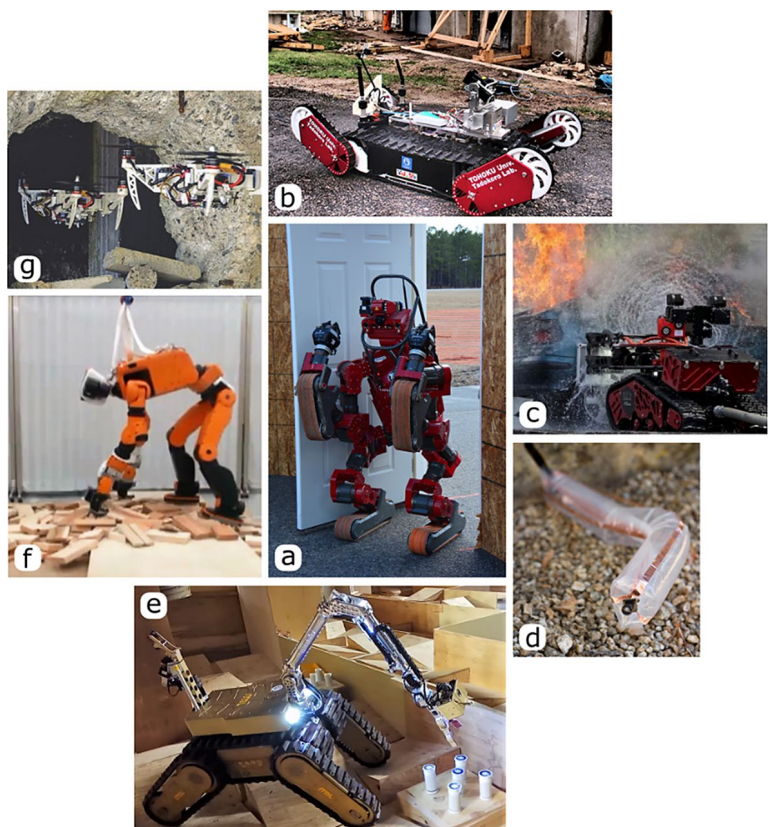

**Fig. 1** Response robots in mission: **a** CHIMP robot performing a DARPA robotic challenge task [14], **b** Quince rescue robot in a disaster site [15], **c** Shark Robotics' Colossus robot extinguishing the fire and clearing away debris [16], **d** Soft growing robot navigating by exploiting the contact with train [17], **e** Karo rescue robot performing dexterity task at 2017 RoboCup competitions [18, 19], **f** Honda's E2-DR walking on scattered debris [20], **g** Quadcopter developed by Swiss roboticists with retracting propeller arms fitting a narrow gap [21]

For instance, a mini-sized underwater robot can agilely accomplish a reconnaissance mission [12, 13], while a maxi-sized tele-operative ground robot is expected to carry out a rubber removal mission effectively.

Since response robots' specifications and applications are vastly diverse, the development and deployment challenges span from mechanical design [22–24] to topics in artificial intelligence (AI) such as robot learning, and planning [25]. As a matter of fact, each aspect of RR is split into several scientific problems that are getting attention in various research communities. Having said that, identifying research trends and the progress made by the researchers in the field of RR requires exacting investigation. Although research works involving implementation with performance evaluation reflect some recent advancements in the field of RR from multiple points of view, it does not delineate a broad picture of the trends and breakthroughs. Review articles are another effort to recognize the research trends in this field [26, 27], however, they are mainly focused on a specific research area of RR and do not offer a comprehensive study over the global research trends.

Bibliometric analysis is an informative tool to study the historical evolution of a scientific field [28], particularly with the rapid growth of publications, which makes analyzing all article challenging. Using bibliometric analysis efficiently reveals several aspects of scientific publications: citation history, the influence of the topic, inter-disciplinarity, the field structure, hotspot subfields, and the distribution layout of contributors (authors, institutions, funding agencies, etc.). Our goal is to identify the research trends of RR in the academic landscape of robotics and to shed light on its recent evolution of research priorities. The central contribution of this work is that the results of our study make it more convenient for researchers to identify the hotspots in RR, especially for the purpose of technical reviews. To that end, this paper analyzes RR literature from multiple bibliometric aspects to realize the research trends and future directions of the field. To that end, we designed our analyses based on the RR-related literature retrieved from the Web of Science, Clarivate Analytics (WOS) database and conducted further data processing and visualization with WOS's analytic tools. We first discussed the implications of overall statistics describing the productivity of authors, publishers, institutions, and countries. Then, we analyzed the contribution of contributing authors in RR in terms of total publication, total citation, and average citation per article. This analysis led to recognizing the leading and influencing scholars in the field. We proceeded with the analysis to investigate the research areas of leading publications and discussed their correlation with the eminent scholars in the field. To realize the clustering of scholars based on their research areas, we investigated the citation network between the contributing authors and extended the analysis to discuss the collaboration between impactful authors in RR. Furthermore, we studied the productivity of journals and conferences publishing RR works in terms of total publications, by which their interest trends were realized. This paper also studies the contribution of productive countries, institutions, and funding agencies from multiple angles such as international collaborations. We examined the timestamped research priorities of RR by conducting a bibliometric analysis on frequently used keywords and technical contents. The findings and implications of the results are also discussed extensively in this paper which reveals the rationale behind the diverse research community of RR. We recognized the most impactful authors, publications, publishers, and productive countries in RR and discussed the possible motivations behind their significant contributions. Motivated by the results of this work, we analyzed the evolution of research priorities of RR and discussed the future possible research trends that concurrently are growing within various domains of AI.

The remainder of this paper is organized as follows: "Methodology and study design" section discusses the bibliometric methodology employed in this paper. In "Results" section, the results of the bibliometric analysis are presented from multiple angles. "Discussion" section develops a discussion on the analysis of the results and their technical and academic implications followed by a conclusion in "Conclusion" section.

## Methodology and study design

There exist many tools and sources of data acquisition, depending on the field of research, to conduct a bibliometric analysis. We designed our analyses based on the insights presented in [29] to select database sources and tools for efficient data processing and visualization. We retrieved the literature related to RR from WOS. Additionally, we employed two specific WOS analytics tools: InCites and Journal Citation Reports (JCR), to carry out citation-based research on sources, institutions, countries, and funding agencies, which provided a supplementary intuition into the prestige of RR literature in the field of robotics.

Concerning the analysis tools, we utilized two software to examine the retrieved data: Bibliometrix R-package [30] and VOSviewer [31]. Bibliometrix was our primary quantitative research tool since it includes more extensive bibliometric analysis techniques. Throughout this work, we combined statistical results of Bibliometrix with the supplementary synthesis obtained from InCites and JCR. VOSviewer was also used to identify the citation network between authors (see "Mapping scientific collaboration" section) and the dynamic of frequent keywords used in RR (see "Analysis of keywords and technical content" section).

To start the research framework, we defined multiple stages for the process. Figure 2 presents the stages of our research methodology and their relationship in this study. We iterated the process between the study design and methodology analysis to achieve a suitable keyword combination; and centralized the study on top authors, publications, sources, organizations, and countries. Then, we applied the content analysis on the outcome and investigated the scientific collaboration between distinguished authors and the most frequent keywords they used in their publications. The outcome of each analysis enabled us to draw conclusions about the global trend in RR research (see "Discussion" section).

During the initial stages, we identified 107 popular keywords after extensively reviewing various types of literature in RR and then classified them into six different categories. Table 1 presents these categories with some of their selected keywords. Clearly, there are numerous connections between all the keywords, searching every combination of them on databases certainly results in RR-related literature. However, not all of them can be a proper search criterion. To obtain the most effective search keyword combination, we iterated our search on WOS using various combinations of keywords, analyzed the outputs, and obtained the parent keywords (e.g., 'disaster robotics' includes all the keywords in Type). This resulted in identifying a combination of 15 comprehensive keywords across all classifications, which can be formulated into WOS search input: "field robotics" OR "rescue robot" OR "disaster robotics" OR "search and rescue robot" OR "response robot" OR "emergency robot" OR "disaster response robot" OR "response robot" OR "urban search and rescue robot" OR "USAR robot" OR "disaster robot" OR "field robot" OR "rescue robotics" OR "response robotics" OR "emergency robotics".

Using the search formula, we adopted "Topic" search method in WOS (January 18, 2021), which detects any document with the conditions of the search criteria in its title, abstract, author keywords, and keywords plus (i.e., words or phrases that frequently appear in the titles of an article's references) and resulted in 1341 documents. In the first attempts of collection analysis, we observed some irrelevant documents in the outputs, studies that are within the field of robotics but do not quite fit the aim of our research (i.e., response robots). For example, roboticists who work in the field of agriculture, commonly use the keyword "field robotics/robot" in their publications that are subcategories of field robotics. Therefore, after several trials of output evaluation, we excluded unrelated documents with a similar approach to excluding 'field robots in agriculture'. Additionally, we omitted duplicated publications due to misspelled or incomplete names of authors. Finally, the purified search yield 1211 documents with 17,412 references, which were from 630 sources, and 3043 authors who contributed to RR between 1991 and 2020.

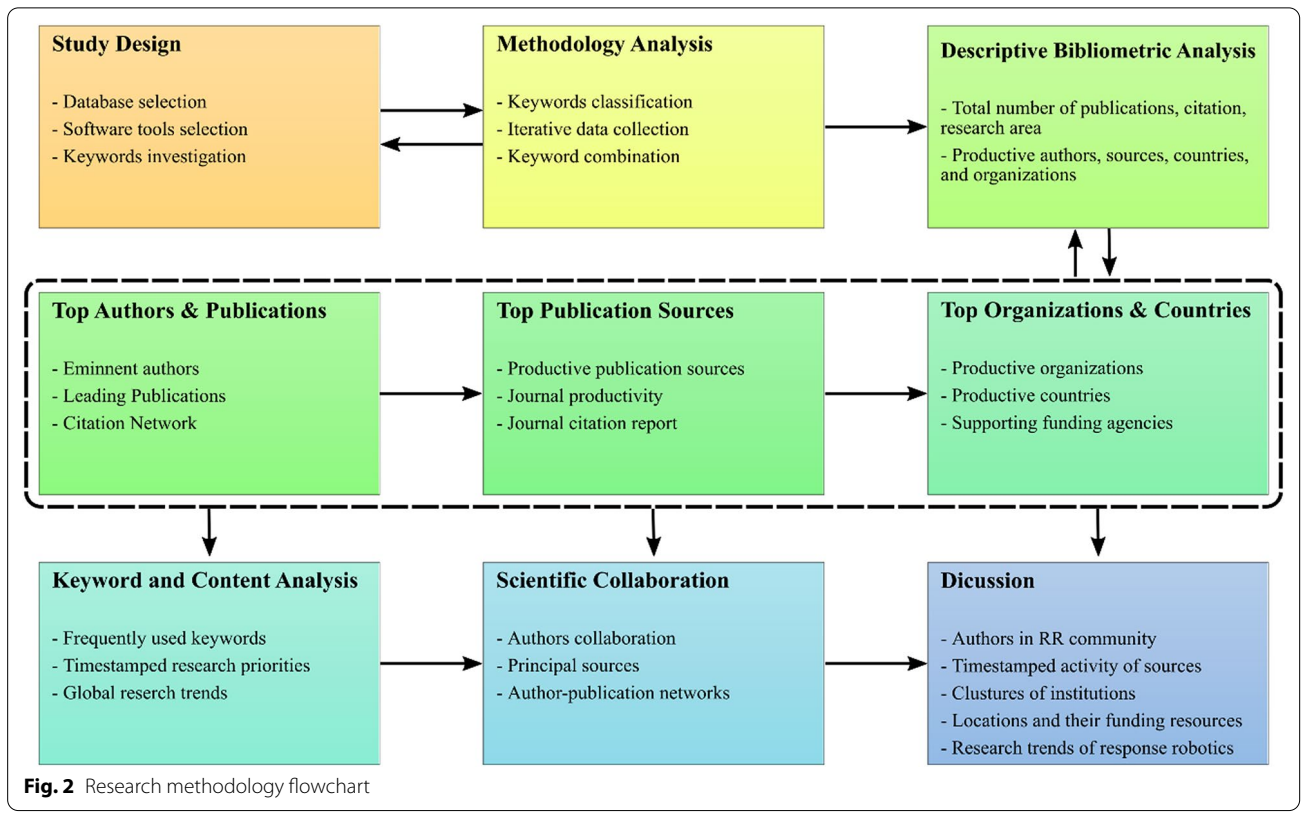

**Fig. 2** Research methodology flowchart

**Table 1** Popular keywords used by authors in RR literature

| Classification | Selected keywords |
| --- | --- |
| Type | UGV, Aquatic, UAV, Aerial, ROV |
| Application | Nuclear, Earthquake, Mine, Firefighting, Medical, Reconnaissance, Emergency response, Tele-presence, Unmanned vehicle, Mobile manipulation, Building and construction, Bomb detection |
| Analysis | Motion analysis, Dynamic analysis, Kinematic analysis, Control architecture, Motion optimization, Inverse kinematics, Object manipulation, Grasping, Stair climbing, Stability, Maneuvering, Mobility, Dexterity, Exploration, Kinetic analysis |
| Hardware | Mechanism design, Robot locomotion, Wheeled robot, Legged robot, Tracked robot, Hybrid locomotion, Manufacturing, Power transmission, Stress analysis, Payload, Mechanical design, Robot arm, Joint mechanism, Prototyping, Passive mechanism, Chassis |
| Software | Obstacle avoidance, Autonomous navigation, SLAM, Object recognition, Vision, Mapping, Exploration, Localization, Motion planning, Path planning, Robot Learning, Motion control, Feature extraction, Path finding, Robot Operating System, Sensor, Algorithm, Autonomy, Simultaneous localization, and mapping, Autonomous, User interface, Human–robot interface, Simulation, Inspection, Perception, Navigation |
| Performance | Field performance, Standard test methods, Prototypical test methods, Performance metrics, Test, RoboCup, Field exercises |

## Results

### Descriptive bibliometric analysis

The refined documents had major document types of 418 articles, 772 proceedings papers, and 21 books. According to the WOS' statistics records, the aforementioned documents are affiliated with 793 institutions from 57 countries. Table 2 lists the frequency of RR-related publications based on WOS records from 1991 to 2020. It also incorporates four citation metrics: total citation, mean total citation per article, mean total citation per year, and citable years received during each year.

The statistics presented in Table 2 is an abstract perspective on RR publications throughout the past decades. The concept of RR started in the early nineteenth century when William L. Whittaker highlighted the evolution of field robots and their necessity in applications such as construction, subsea, space, nuclear, mining, and military applications. In the late nineteenth century, this field received more attention once Japanese scientists such as Kennichi Tokuda and Masayuki Nunobiki introduced practical applications of response robots in rescue missions. Research in RR expanded more in the following years. Based on WOS records, 2016 and 2017 were among the most productive years with 102 and 109 publications, respectively. The significant total citations of 1752 in 2013 indicate the influence of RR publications among roboticists.

Publication records associated with RR on WOS are categorized in many research areas that often one document overlaps multiple areas. Robotics with 1061,

**Table 2** Statistics for literature published related to RR

| Year | Number of publications | Cumulative number of publications | Total citation | Mean total citation per article | Mean total citation per year |
| --- | --- | --- | --- | --- | --- |
| 1991 | 1 | 1 | 0 | 0 | 0 |
| 1992 | 0 | 1 | 0 | 0 | 0 |
| 1993 | 1 | 2 | 0 | 0 | 0 |
| 1994 | 0 | 2 | 0 | 0 | 0 |
| 1995 | 1 | 3 | 7 | 7.000 | 0.269 |
| 1996 | 3 | 6 | 34 | 11.333 | 0.453 |
| 1997 | 0 | 6 | 0 | 0.000 | 0.000 |
| 1998 | 3 | 9 | 16 | 5.333 | 0.232 |
| 1999 | 10 | 19 | 67 | 6.700 | 0.305 |
| 2000 | 6 | 25 | 41 | 6.833 | 0.325 |
| 2001 | 4 | 29 | 158 | 39.500 | 1.975 |
| 2002 | 21 | 50 | 155 | 7.381 | 0.388 |
| 2003 | 19 | 69 | 298 | 15.684 | 0.871 |
| 2004 | 23 | 92 | 643 | 27.957 | 1.645 |
| 2005 | 49 | 141 | 503 | 10.265 | 0.642 |
| 2006 | 53 | 194 | 380 | 7.170 | 0.478 |
| 2007 | 53 | 247 | 331 | 6.245 | 0.446 |
| 2008 | 68 | 315 | 416 | 6.118 | 0.471 |
| 2009 | 70 | 385 | 856 | 12.229 | 1.019 |
| 2010 | 52 | 437 | 726 | 13.962 | 1.269 |
| 2011 | 53 | 490 | 365 | 6.887 | 0.689 |
| 2012 | 62 | 552 | 365 | 5.887 | 0.654 |
| 2013 | 72 | 624 | 1752 | 24.333 | 3.042 |
| 2014 | 86 | 710 | 517 | 6.012 | 0.859 |
| 2015 | 86 | 796 | 365 | 4.244 | 0.707 |
| 2016 | 102 | 898 | 497 | 4.873 | 0.975 |
| 2017 | 109 | 1007 | 436 | 4.000 | 1.000 |
| 2018 | 62 | 1069 | 310 | 5.000 | 1.667 |
| 2019 | 86 | 1155 | 164 | 1.907 | 0.953 |
| 2020 | 56 | 1211 | 26 | 0.464 | 0.464 |

Automation Control Systems with 964, Computer Science with 880, Engineering with 779, and Instrument Instrumentation with 207 record counts are among the most common WOS research areas that we found in our analysis. The most productive eminent authors, listed in the WOS database, who substantially contributed to RR are Satoshi Todokoro with 43, Robin R. Murphy with 32, and Fumitoshi Matsuno with 28 publications. Japan, China, and the USA with 272, 198, and 167 publications are indexed as the most prolific countries in RR literature, respectively. Moreover, our analysis from the WOS data indicated that 43, 40, and 28 publications in RR were recorded from authors with Waseda University, Tohoku University, and Kyoto University, respectively. Further analysis regarding the most influential authors, organizations, sources, countries, and funding agencies are presented in "Eminent authors and leading publications", "Impactful and productive publication sources" and "Contribution of organizations and locations" sections.

### Eminent authors and leading publications
*Eminent authors*

This section discusses the contributions of authors to RR to recognize leading and impactful scholars in the field. Table 3 presents the top 14 leading authors in RR who have the highest number of publications and citations. Tadokoro who ranks first in terms of total publications has conducted comprehensive research on RR and rescue robots during his active years (since 2000, according to the selected publications). He participated in the rescue robot Quince project employed for the investigation of the Fukushima Daiichi nuclear power plant damaged by a big tsunami [15]. Moreover, he has actively contributed to RoboCup Rescue Robot League and has published several articles on this matter [32]. Murphy is another eminent scholar who has contributed to RR significantly by exploring various ways of deploying rescue robots into disaster sites [33]. On the subject of RR, Murphy's research includes the application of multi-robot systems in rescue operations [34], human–robot interaction in urban search and rescue (USAR) [35], developments of disaster robotics through USAR competitions [36], and assessment of NIST standard test methods for response robots [37]. According to the results presented in Table 3, Murphy's research has attracted the highest attention among all 14 leading authors of the community in terms of total citations. From the ACPP point of view, Jacoff's publications maintain a high citation per publication ratio which indicates the significance of his conducted research. Jacoff's research predominantly focuses on the test and evaluation of response robots by developing standard test methods under NIST [38]. Inspired by the frameworks and measures developed as NIST standard test methods for response robots, the Rescue RoboCup competitions have played a major role in encouraging design, development, and evolvement of rescue robots during the past 2 decades under Jacoff's supervision [39].

Although the authors mentioned in Table 3 are specifically recognized for their significant contribution to RR, some of them are also widely well-known for their influential research in robotics and AI in general. For instance, Siegwart is a globally significant scholar in the field of mobile robotics while has contributed to several research projects in RR from multiple aspects such as aerial-ground collaboration, and multiagent systems [40, 41]. These active contributions of leading scholars from

**Table 3** Contribution of eminent authors in RR research

| Rank | Author | TP | TC | ACPP | Institution |
| --- | --- | --- | --- | --- | --- |
| 1 | Tadokoro S | 43 | 76 | 1.77 | Tohoku University |
| 2 | Murphy R | 32 | 402 | 12.56 | Texas A&M University |
| 3 | Matsuno F | 28 | 31 | 1.11 | Kyoto University |
| 4 | Birk A | 23 | 70 | 3.04 | Jacobs University |
| 5 | Ohno K | 13 | 39 | 3.00 | Tohoku University |
| 6 | Amano H | 12 | 19 | 1.58 | National Research Institute of Fire and Disaster |
| 7 | Hirose S | 12 | 88 | 7.33 | Hibot Corp. |
| 8 | Ito N | 11 | 29 | 2.64 | Aichi Institute of Technology |
| 9 | Kamegawa T | 11 | 48 | 4.36 | Okayama University |
| 10 | Montambault S | 9 | 29 | 3.22 | Hydro-Quebec IREQ |
| 11 | Tsukagoshi H | 9 | 41 | 4.56 | Tokyo Institute of Technology |
| 12 | Jacoff A | 8 | 84 | 10.50 | National Institute of Science and Technology |
| 13 | Pouliot N | 8 | 21 | 2.63 | Hydro-Quebec IREQ |
| 14 | Siegwart R | 8 | 22 | 2.75 | ETH Zurich |

*TP* total publications; *TC* total citations; *ACPP* average citation per publication a wide range of disciplines to RR emphasize the importance and multidisciplinary nature of the field.

### *Leading publications*

The multidisciplinary nature of RR is recognizably reflected in Table 4 summarizing the top 15 leading publications in the field. The leading publications listed in Table 4 are roughly distributed over various topics in science and engineering. For instance, [42] investigates the human–robot interaction problems in RR and proposes a preliminary domain theory of the visual technical search task. While Murphy's work addresses problems in RR from a computer science perspective, Kamegawa discusses mobility and maneuvering problems by presenting a snake-like mechanism for rescue robots in [43]. Rescue robots' mechanisms and maneuvering problems have also been studied by authors conducting research on novel and state-of-art topics such as soft robotics [44, 45]. On this subject, Rich reviews untethered soft mechanisms to be deployed as wearable robots with applications in field robotics [46]. Moreover, there are some mutual research problems and gaps between RR and mobile robotics, which have been addressed by a couple of works listed in Table 4 accordingly, e.g., Smith's research on outdoor navigation and mapping datasets for field robots [47]. Altogether, the results confirm that the diversity of challenges for the development and deployment of rescue robots have been addressed by a broad range of publications from various disciplines.

Drawing a comparison between the results presented in Tables 3 and 4 reveals important bibliometric aspects of RR. Among the top 15 leading publications listed in Table 4, there are only two publications authored by eminent scholars in RR. In other words, leading publications in RR are not necessarily produced by leading authors in the field. This fact plausibly implies that RR cannot be recognized as a closed research community since it has received extensive attention from broad disciplines in science and engineering. Besides, the overlap between eminent authors listed in Table 3 and the authors' published works mentioned in Table 4 recognizes the most impactful scholars in the field, namely Murphy and Kamegawa.

### *Citation network*

Although recognizing eminent authors and leading publications insightfully highlights impactful people and works in RR, it does not capture collaboration and the relationship between authors and their works. This relationship potentially can be analyzed by the way of different factors such as co-authorship, co-occurrence, bibliometric coupling, and citation according to VOSviewer categories. We proceeded with the analysis with the citation factor since it reveals broader relationships between the authors in the field and facilitates clustering the authors for further discussions. This analysis has been performed using tools in VOSviewer to visualize the citation network of the authors included in the bibliometric dataset, as Fig. 3 depicts the results.

Figure 3 preliminary demonstrates the impact of the eminent authors of RR discussed in "Eminent authors" section. For instance, the citation network clearly emphasizes the significant academic attention that Murphy, Tadokoro, Matsuno, and Brik have received as impactful scholars in the field. Besides, the way that the authors have been clustered reveals supplementary

**Table 4** Top 15 leading publications in RR

| Paper | TC | TCPY |
| --- | --- | --- |
| Murphy [42], IEEE Transactions on Systems, Man, and Cybernetics: Part C | 256 | 14.22 |
| Rich et al. [46], Nature Electronics | 199 | 49.75 |
| Leonard et al. [48], Journal of Field Robotics | 197 | 16.41 |
| Smith et al. [47], The International Journal of Robotics Research | 173 | 13.30 |
| Murphy et al. [33], IEEE Robotics & Automation Magazine | 111 | 8.53 |
| Kamegawa et al. [43], IEEE International Conference on Robotics and Automation (ICRA) | 106 | 5.88 |
| Burdick and Fiorini [49], The International Journal of Robotics Research | 104 | 5.47 |
| Hollinger and Sukhatme [50], The International Journal of Robotics Research | 102 | 12.75 |
| Poppinga et al. [51], 2008 IEEE/RSJ International Conference On Robots And Intelligent Systems | 96 | 6.85 |
| Liu and Nejat [52], Journal of Intelligent & Robotic Systems | 95 | 10.55 |
| Spagna et al. [53], Bioinspiration & Biomimetics | 95 | 6.33 |
| Pathak et al. [54], Journal of Field Robotics | 71 | 5.91 |
| Hygounenc et al. [55], The International Journal of Robotics Research | 69 | 3.83 |
| Kawatsuma et al. [56], Industrial Robot | 65 | 6.50 |

*TCPY* total citation per year

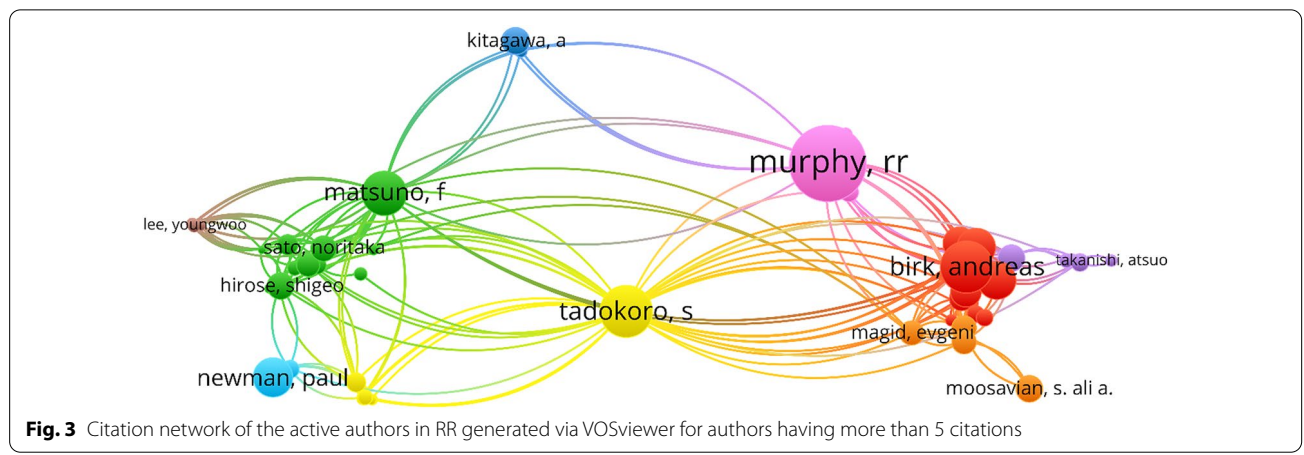

**Fig. 3** Citation network of the active authors in RR generated via VOSviewer for authors having more than 5 citations

information about their mutual sub-fields in RR. According to the carried-out analysis, 87 authors having more than five citations have been grouped into nine clusters in the network. In fact, the clustering in the citation network not only reflects the collaboration and co-authorship between the in-network authors but also adequately explains how the research works of those authors relate to each other, as manifests in cluster 5. This cluster, shown in purple in Fig. 3, includes Jacoff, Pellenz, and Suthakorn who have published multiple articles addressing performance evaluation and standard test method development for rescue robots [57]. On the other hand, authors grouped in cluster 1, shown in red in the figure, have contributed to RR from AI and computer science perspectives, namely Nejat [58], Brik [59], and Schwertfeger [60]. Moreover, the clustering of the in-network authors has captured the nationality as well as the mutual sub-fields of research and co-authorships. Cluster 2, shown in green, only contains authors from Japan who have been focusing on RR from the mechanical engineering point of view, namely Matsuno [61], Hirose [62], and Gofuku [63].

### Impactful and productive publication sources

This section presents the top contributing sources in the field of RR and their production rates. The following subsections centralize the analysis on their productivity in RR; and examines the influence and contribution of the top 5 leading journals based on the outcomes of InCites and JCR in the field of robotics.

#### *Top 20 sources*

In this part, we focus on analyzing journals and conference proceedings as the primary sources of publications and their share in broadcasting RR-related research. Based on this frame, there are 418 research articles published in journals, and 772 proceedings papers published in conferences. The top 15 journals that published articles related to RR are ranked in Table 5. In this list, *Advanced Robotics* is the leading journal with 52 totals; and *Journal of Field Robotics*, *Journal of Robotics and Mechatronics*, *International Journal of Advanced Robotic Systems*, and *International Journal of Robotics Research* take the second to fifth positions, respectively. The preceding journals approximately hold the share of 41% of the total RR journal research articles, and *Advanced Robotics* exclusively owns about 14%. The last four rankings in the list, only published about 6% of RR articles during 1991–2020, which indicates their concentration on different topics in robotics. Note that the first and the last journal in the list have a gap of 48 publications. Additionally, *Springer* is the most contributing publisher with 4 journals in the top 15 list.

**Table 5** Top 15 journals published research articles related to RR (1991–2020)

| Rank | Journals | Publisher | Articles | Impact factor |
|---|---|---|---|---|
| 1 | Advanced Robotics | Taylor & Francis | 52 | 1.69 |
| 2 | Journal of Field Robotics | Wiley | 37 | 3.76 |
| 3 | Journal of Robotics and Mechatronics | Fuji Technology Press | 26 | 0.89 |
| 4 | International Journal of Advanced Robotic Systems | Sage Publications | 19 | 1.65 |
| 5 | International Journal of Robotics Research | Sage Publications | 19 | 4.70 |
| 6 | Industrial Robot | Emerald Publishing | 17 | 1.12 |
| 7 | Sensors | MDPI | 12 | 3.57 |
| 8 | Journal of Intelligent and Robotic Systems | Springer | 11 | 2.64 |
| 9 | Robotica | Cambridge University Press | 11 | 2.08 |
| 10 | Robotics and Autonomous Systems | Elsevier | 11 | 3.12 |
| 11 | IEEE Robotics and Automation Magazine | IEEE | 10 | 5.14 |
| 12 | Autonomous Robots | Springer | 7 | 3.00 |
| 13 | Robotics | MDPI | 7 | 2.94 |
| 14 | Artificial Life and Robotics | Springer | 6 | 0.81 |
| 15 | Intelligent Service Robotics | Springer | 4 | 2.24 |

The conferences in the field of robotics are generally favorable between RR researchers and usually highly competitive to publish. Hence, we included their ranking to show conference proceedings contributions as well. Similarly, Table 6 lists the top 15 productive conference proceedings in the field of RR. Note that Table 6 presents the conferences based on their proceedings in each year, not the accumulated published papers. The *Conference of the Society of Instrument and Control Engineers of Japan* (*SICE*) with 13 papers in 2008 has the lead in publications per year. It is followed by *IEEE International Workshop on Safety Security and Rescue Robotics* (*SSRR*), *IEEE/RSJ International Conference on Intelligent Robots and Systems* (*IROS*), and *IEEE International Conference on Robotics and Biomimetics* (*ROBIO*) in the list.

Overall, conferences such as *SSRR* with 41 papers, *IROS* with 20 papers, *ICRA* with 11 papers, and *SICE* with 20 papers, were highly productive in several years. Although the first ranks (*Advanced Robotics* and *SICE*) in the top lists have a considerable difference in the total number of publications, it is complicated to make comparisons between sources because Table 6 only illustrates year publications. However, the decrease in the total number of RR papers after 2008 is notable among conferences. In the next section, we discuss the productivity of the top journals from the time point, which may be compared to the information in Table 6.

### Journal productivity

The attention of researchers to a particular field is inherently dynamic and depends on a variety of factors (e.g., research gaps, funding availability, industry demands, etc.). The field of RR follows similar alternations. Disastrous situations such as USAR [2], nuclear field emergency operations [3], mine rescue missions [4], or earthquake crisis management [6] accentuated getting assistant from robots where it is highly unsafe for humans. The aforementioned causes are observable from the trends of publications in the historical map presented in Fig. 4. For instance, the Great Hanshin Earthquake or Kobe earthquake motivated scholars in Kobe University in Japan to initiate early generations of RR research and this topic became favorable among researchers in the late 2000s. *Advanced Robotics* journal has been consistently productive in publishing research works in this field since the late 90s. With almost a ten-year gap *Journal of Field Robotics* and *Journal of robotics and mechatronics* started a similar production rate. The *Robotics* and *Sensors* journal, on the other hand, expressed interest in response robots between 2010 and 2015. Furthermore, scientific competitions such as DARPA Grand Challenge or RoboCup Competitions pursued by funding agencies support in the early 2000s expedited the publications.

**Table 6** Top 15 conference proceedings published articles related to RR (1991–2020)

| Rank | Conferences | Year | Papers |
| --- | --- | --- | --- |
| 1 | Conference of the Society of Instrument and Control Engineers of Japan | 2008 | 13 |
| 2 | IEEE International Workshop on Safety Security and Rescue Robotics | 2005 | 11 |
| 3 | IEEE International Workshop on Safety Security and Rescue Robotics | 2007 | 9 |
| 4 | IEEE/RSJ International Conference on Intelligent Robots and Systems | 2006 | 8 |
| 5 | IEEE International Conference on Robotics and Biomimetics | 2008 | 8 |
| 6 | IEEE International Workshop on Safety Security and Rescue Robotics | 2012 | 8 |
| 7 | Conference of the Society of Instrument and Control Engineers of Japan | 2006 | 7 |
| 8 | IEEE International Workshop on Safety Security and Rescue Robotics | 2013 | 7 |
| 9 | IEEE International Conference on Robotics and Automation | 2004 | 6 |
| 10 | IEEE/RSJ International Conference on Intelligent Robots and Systems | 2007 | 6 |
| 11 | International Conference on Applied Robotics for the Power Industry | 2010 | 6 |
| 12 | International Conference on Ubiquitous Robots and Ambient Intelligence | 2017 | 6 |
| 13 | IEEE International Workshop on Safety Security and Rescue Robotics | 2017 | 6 |
| 14 | IEEE/RSJ International Conference on Intelligent Robots and Systems | 2003 | 6 |
| 15 | IEEE International Conference on Robotics and Automation | 2005 | 5 |

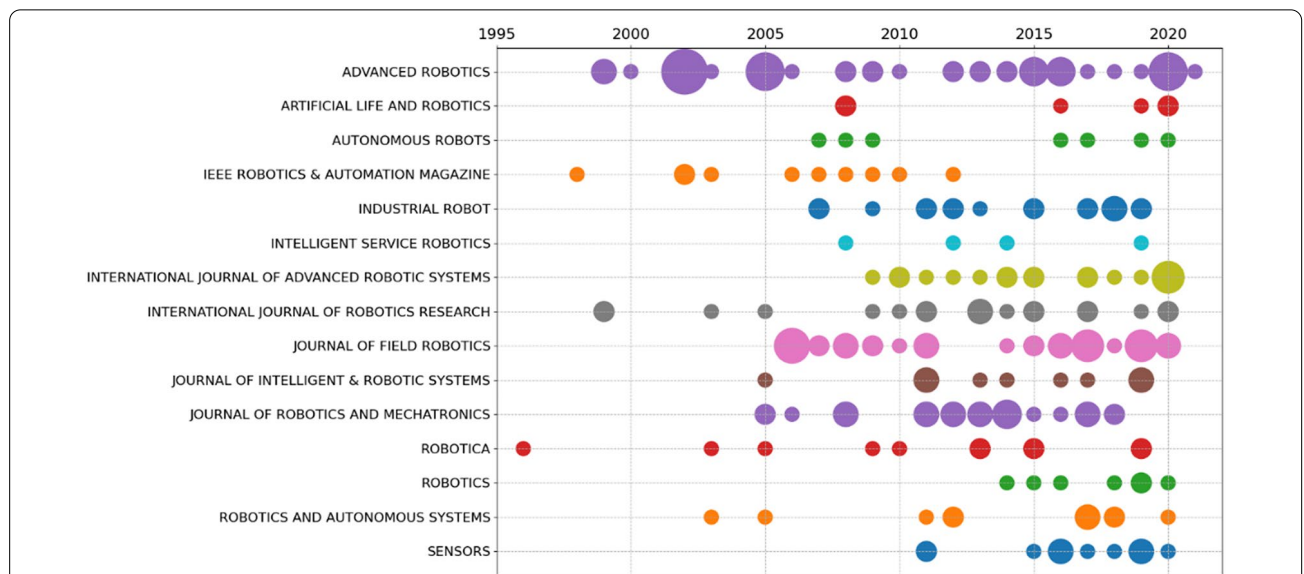

**Fig. 4** Productivity of top 15 journals published RR articles in 1995–2020

The previous analysis of journals was based on the number of RR articles published. From a higher perspective, Table 7 presents supplementary data extracted from JCR of the top 5 journals in the field of robotics ("Robotics" category was selected as the "Research Area" in WOS). This table provides more insight into the leadership of these journals in robotics besides their contribution to RR. According to WOS results, the *International Journal of Robotics Research* and *Journal of Field Robotics*, are ranked among the top 5 journals (see Table 5); they have the highest impact factor and article influence in robotics, while one has received the highest and the other, the lowest total citations for robotics-related publications. The other three journals in Table 7, even though they are ranked 8–10 in Table 5, they have a considerable overall contribution to the field of robotics.

Journal evaluation metrics are essentially correlated, and it is useful to compare them from various perspectives. Figure 5 depicts the share of each of the top 5 journals from the total robotics papers on WOS. From this standpoint, all the aforementioned journals (except *Journal of Field Robotics*) hold major shares (~20%) of total publications in the field of robotics. This highlights the significance of the RR topic among researchers that highly impactful and productive journals in robotics actively publish related articles.

### Contribution of organizations and locations

In this section, we analyze institutions and locations that are affiliated with RR research, and we go beyond the number of publications. The citation-based research analytics tool of WOS, InCites, was used to extract complementary metrics to the initial Bibliometrix data.

**Table 7** Journal Citation Reports-InCites: top 5 cited journals published robotics articles (1991–2020) according to WOS' JCR in 2019

| Rank | Journals | TC | TD | AI | IF |
| --- | --- | --- | --- | --- | --- |
| 1 | International Journal of Robotics Research | 99,070 | 2253 | 1.72 | 4.703 |
| 2 | Robotics and Autonomous Systems | 65,850 | 3014 | 0.744 | 2.825 |
| 3 | Journal of Intelligent and Robotic Systems | 32,269 | 2899 | 0.472 | 2.259 |
| 4 | Robotica | 27,345 | 2677 | 0.281 | 1.509 |
| 5 | Journal of Field Robotics | 23,181 | 859 | 1.059 | 3.581 |

*TC* times cited; *TD* total WOS documents; *AI* article influence; *IF* impact factor

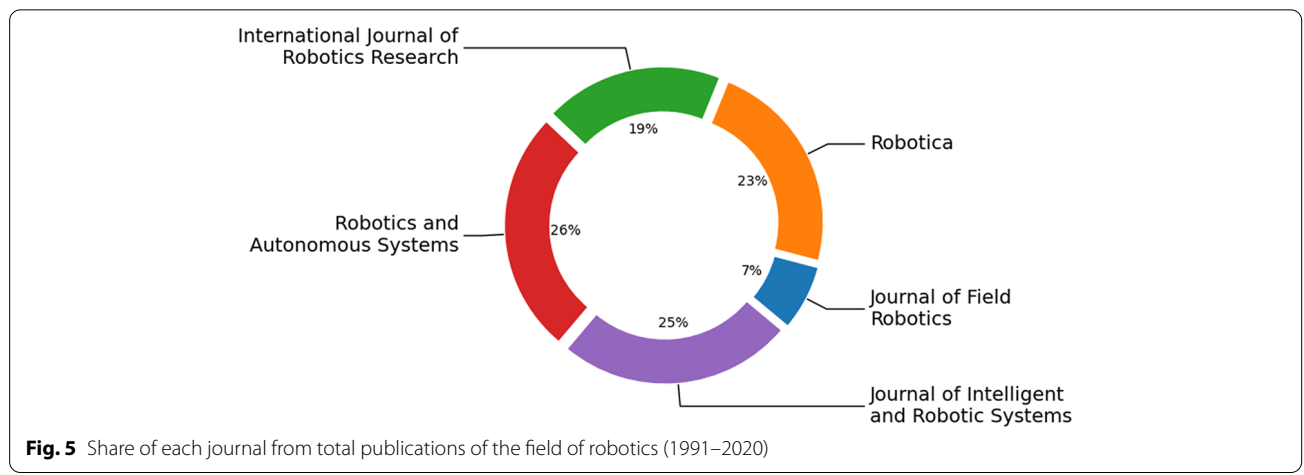

**Fig. 5** Share of each journal from total publications of the field of robotics (1991–2020)

Furthermore, we present the top funding agencies that have supported research in robotics across the globe.

### Institutions

The affiliation of researchers is a major part of the bibliometric analysis and in this study, authors who published articles concerning RR are affiliated with 793 institutions (only 29 universities have ≥ 10 publications). As mentioned earlier, RR is a multidisciplinary field that demands the collaboration of authors from various backgrounds. Institutions are a primary source of academic research that gather researchers and carry this responsibility. Here, we present institutions that have remarkable interest in RR based on the number of articles published by authors with their affiliation (Table 8). Waseda University holds first place among the institutions with 43 published articles in RR. Institute for Disaster Response Robotics at Waseda University with the principal research on "Four-arm, four-crawler disaster response robot" and "Legged robot with high locomotion and manipulation ability," evidently represents the highest number of publications. The Human–Robot Informatics Laboratory positions Tohoku University in the second place with a very close distance to Waseda University. One particular insight from Table 8 is that all first four or in fact, eight out of the first 15 most contributing organizations are located in Japan. This indicates the particular

**Table 8** Top 15 institutions contributed to RR along with WOS report metrics for the Robotics research area (1991–2020)

| Affiliations | Year | Articles | WOS report (InCites) | | | | |
|---|---|---|---|---|---|---|---|
| | | | Rank* | TD | TC | % DC | CI |
| Waseda University | 2006–2020 | 43 | 102 | 1134 | 5066 | 64.20 | 0.87 |
| Tohoku University | 2006–2020 | 40 | 65 | 1056 | 7654 | 71.97 | 1.07 |
| Kyoto University | 1999–2020 | 28 | 89 | 668 | 5567 | 65.42 | 0.93 |
| Tokyo Institute of Technology | 2002–2020 | 28 | 46 | 1099 | 9350 | 70.06 | 1.20 |
| University of Sydney | 2006–2020 | 27 | 26 | 606 | 14,360 | 81.19 | 2.50 |
| Carnegie Mellon University | 1991–2020 | 23 | 1 | 2586 | 55,814 | 80.94 | 2.83 |
| University of Ulsan | 2008–2015 | 20 | 434 | 134 | 1137 | 55.22 | 0.84 |
| Texas A&M University | 2011–2020 | 19 | 109 | 501 | 4790 | 68.46 | 1.53 |
| University of Toronto | 2010–2019 | 19 | 56 | 791 | 8325 | 77.62 | 1.39 |
| University of Electro-Communication | 2005–2015 | 18 | 132 | 675 | 4073 | 65.19 | 0.91 |
| Okayama University | 2004–2020 | 17 | 228 | 430 | 2270 | 61.16 | 0.91 |
| University of Tokyo | 1999–2020 | 15 | 10 | 2471 | 23,546 | 70.62 | 1.39 |
| China University of Mining and Technology | 2009–2020 | 14 | 525 | 178 | 854 | 64.04 | 0.67 |
| Jacobs University Bremen | 2006–2019 | 14 | 359 | 141 | 1431 | 75.89 | 1.43 |
| Nagoya Institute of Technology | 2011–2019 | 14 | 450 | 275 | 1086 | 56.73 | 0.82 |

*TD* total WOS documents; *TC* total citation of WOS documents; *% DC* percent of WOS documents cited; *CI* citation impact focus of robotics research in this country on RR, which is discussed thoroughly in this work.

InCites was used to produce additional citation-based research analytics metrics. We selected the "Robotics" research area in InCites and extracted the overall performance of each institution between 1991 and 2020. An interesting observation from this WOS data is that well-known institutions such as Carnegie Mellon University, University of Tokyo, and the University of Sydney have significant impacts on the field of robotics, although their primary interest is not RR. The University of Tokyo is the only Japanese institution in the list that its comprehensive performance places it next to CMU and USYD on top of the WOS ranking. However, the percent of WOS documents cited from all the organizations listed in Table 8 is pretty close, which represents their impactful participation in the field of robotics.

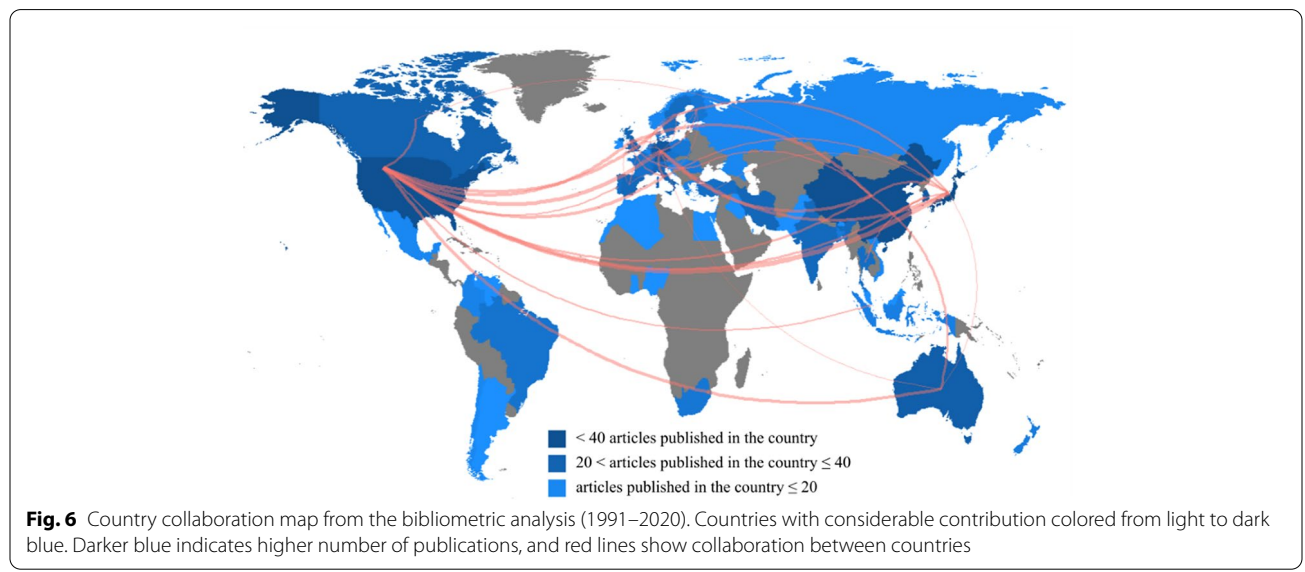

**Fig. 6** Country collaboration map from the bibliometric analysis (1991–2020). Countries with considerable contribution colored from light to dark blue. Darker blue indicates higher number of publications, and red lines show collaboration between countries

**Table 9** Top 15 countries contributed to RR (1991–2020)

| Rank | Country | Articles | Freq. % | SCP | MCP | MCP ratio | WOS report InCites | |
|---|---|---|---|---|---|---|---|---|
| | | | | | | | TD | % DC |
| 1 | Japan | 272 | 23.15 | 264 | 8 | 0.0294 | 24,674 | 61.21 |
| 2 | China | 197 | 16.85 | 173 | 24 | 0.1218 | 32,035 | 53.65 |
| 3 | USA | 167 | 14.21 | 143 | 24 | 0.1437 | 35,267 | 73.48 |
| 4 | Korea | 78 | 6.64 | 77 | 1 | 0.0128 | 8801 | 61.54 |
| 5 | Germany | 74 | 6.30 | 61 | 13 | 0.1757 | 12,697 | 71.39 |
| 6 | Australia | 49 | 4.17 | 40 | 9 | 0.1837 | 4294 | 75.01 |
| 7 | Canada | 37 | 3.15 | 35 | 2 | 0.0541 | 6961 | 73.71 |
| 8 | Iran | 28 | 2.38 | 26 | 2 | 0.0714 | 2528 | 61.16 |
| 9 | Italy | 23 | 2.04 | 20 | 3 | 0.1304 | 8752 | 72.25 |
| 10 | India | 23 | 1.96 | 23 | 0 | 0 | 3422 | 54.70 |
| 11 | Spain | 22 | 1.87 | 20 | 2 | 0.0909 | 5908 | 71.80 |
| 12 | UK | 22 | 1.87 | 18 | 4 | 0.1818 | 8668 | 69.92 |
| 13 | Thailand | 16 | 1.36 | 14 | 2 | 0.125 | 868 | 46.31 |
| 14 | Portugal | 13 | 1.11 | 10 | 3 | 0.2308 | 2171 | 70.47 |
| 15 | Turkey | 13 | 1.11 | 12 | 1 | 0.0769 | 1374 | 69.72 |

*SCP* single country publications; *MCP* multiple country publications; *MCP Ratio* MCP/SCP; *TD* WOS documents; *% DC* percent of WOS documents cited

*Countries*

The retrieved literature related to RR are originated from 57 countries. Figure 6 presents the collaboration world map and the contribution of each country. Since it is not possible to assume all the 57 countries that directly contributed to this field, we ranked the top 15 contributing countries in Table 9. The data show that only 12 countries published more than 20 documents in 1991–2020. Hence, the rest of the listed countries may be enumerated due to international collaborations or authors with more than one affiliation. Table 9 also shows that Japan with a considerable gap in the total number of published documents and the frequency of publications stands in the first place. This confirms the previous analysis about Japanese institutions that were in the top list of contributors to RR. Japan's publications are substantially submitted by researchers within the country (low MPC ratio). However, Korea and India both have lower MCP ratios compared to Japan, this metric is more significant for Japan because of its high contributions.

The SCP of Japan indicates the special interest of Japanese researchers in RR and perhaps a global base for response robot scholars in the world. China and the USA follow Japan with substantial numbers of RR-related publications, the USA with 35,267 publications (1991–2020) in the field of robotics also holds the overall first place. Additionally, they have the highest MCPs that represent high international collaborations for both countries. The combination of funding availability, economy, and leading institutions essentially influence the research performance of countries, which is comprehensible for the top countries.

*Funding agencies*

There are complicated correlations between different segments of a research area (e.g., researchers, countries, or institutions) that require several angles to thoroughly analyze the dynamics. In general, the productivity of an author may not regard a particular factor such as his or her special research interest or the support received from the affiliated institution. To be thorough, we accounted

**Table 10** Top 10 international agencies funded institutions of Table 8 for research in robotics based on InCites report (1991–2020)

| Rank | Name | TD | TC | % DC | CI |
|---|---|---|---|---|---|
| 1 | National Science Foundation (NSF) | 475 | 8233 | 80.21 | 2.88 |
| 2 | Japan Society for the Promotion of Science (JSPS) | 696 | 3304 | 63.22 | 0.82 |
| 3 | Ministry of Education, Culture, Sports, Science and Technology, Japan (MEXT) | 552 | 3050 | 71.92 | 0.94 |
| 4 | Australian Research Council (ARC) | 98 | 2317 | 86.73 | 3.91 |
| 5 | Natural Sciences and Engineering Research Council of Canada (NSERC) | 163 | 1578 | 79.75 | 1.55 |
| 6 | Defense Advanced Research Projects Agency (DARPA) | 75 | 1470 | 82.67 | 2.12 |
| 7 | Office of Naval Research (ONR) | 112 | 1352 | 75.00 | 2.04 |
| 8 | Kakenhi | 398 | 1254 | 59.55 | 0.72 |
| 9 | US Army Research Lab (ARL) | 44 | 1098 | 88.64 | 3.24 |
| 10 | National Natural Science Foundation of China (NSFC) | 183 | 937 | 66.12 | 0.69 |

*TD* WOS documents; *TC* times WOS documents cited; *% DC* percent of WOS documents cited; *CI* citation impact of WOS documents for the influence of funding agencies to fuel the research directions and contributions in robotics. Not to imply that it completely determines the productivity in this field, but it is remarkable enough to encourage researchers to progress in a research domain. Table 10 shows the top 10 international agencies funded research in robotics of the top institutions in RR (Table 10).

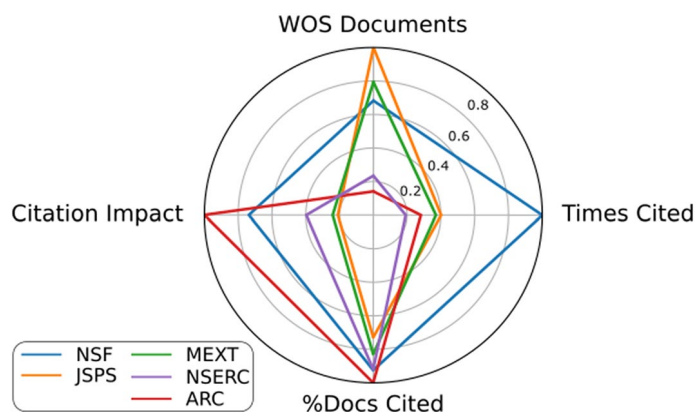

**Fig. 7** Performance of top 5 international agencies funded research in robotics based on InCites report (1991–2020)

An outstanding insight from the above data is that four USA funding agencies (NSF, DARPA, ONR, and ARL) are on the list. NSF, on average, has the highest total WOS documents, total (percent) citation, and citation impact. Figure 7 confirms that NSF has overall superiority, JSPS published the highest WOS documents, and ARC owns the greatest citation statistics (CI and %Docs Cited), proportional to other top 5 agencies. It should be noted that Fig. 7 represents the normalized information of Table 10. Broadly speaking, almost most of the publications (> 60%) of these top funding agencies are cited in other publications. Following the USA, three Japanese funding agencies (JSPS, MEXT, and Kakenhi) are in the list that verifies Japan's ranking (1st in Table 10) between other countries. It is also remarkable that only one Chinese funding agency (NSFC) is among the top robotics research supporting agencies, while China owns second place within the list of most contributing countries (Table 10).

### Analysis of keywords and technical content

Conducting a bibliometric timestamped keyword analysis and investigating the frequently used keywords within their scientific context facilitate realizing the research trends in a field, however, time-independent and timestamped analysis should be performed complementarily to develop a backboned bibliometric discussion. This section discusses both time-independent and timestamped approaches to emphasize the importance of incorporating the time element and to develop an argument on the way that the research priorities have evolved in RR during the last decade. Figure 8 demonstrates the results of the time-independent keyword analysis. As the analysis suggests, various research paradigms in RR have been represented by the frequent keywords mentioned in the figure. For instance, track, snake-like robot, and jumping robot refer to the locomotion mechanisms of response robots, while visual odometry, image processing, user interface, and navigation represent computer-related research topics within the field. Moreover, different variants of response robots made it through the most frequent keyword list such as marine, coal/mine, service, and disaster robots. Nevertheless, the mega keywords (e.g., rescue robots, field robotics, mobile robotics, and navigation) mostly refer to generic topics in RR from which no discussion can be derived about the time-factored research proprieties and trends in the field.

To provide a complementary analysis, the frequent keywords have been considered their time context as Fig. 9 demonstrates. According to the results, three phases with distinct research priorities can be recognized in RR during the past decade: (1) fundamental challenges, (2) control and robotics challenges, and (3) AI and machine learning challenges.

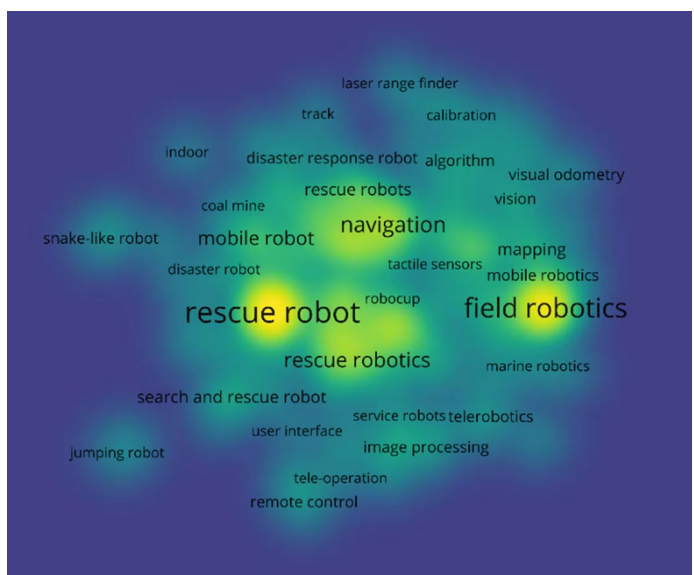

**Fig. 8** Frequent keywords RR literature (brighter colors reflect more frequency)

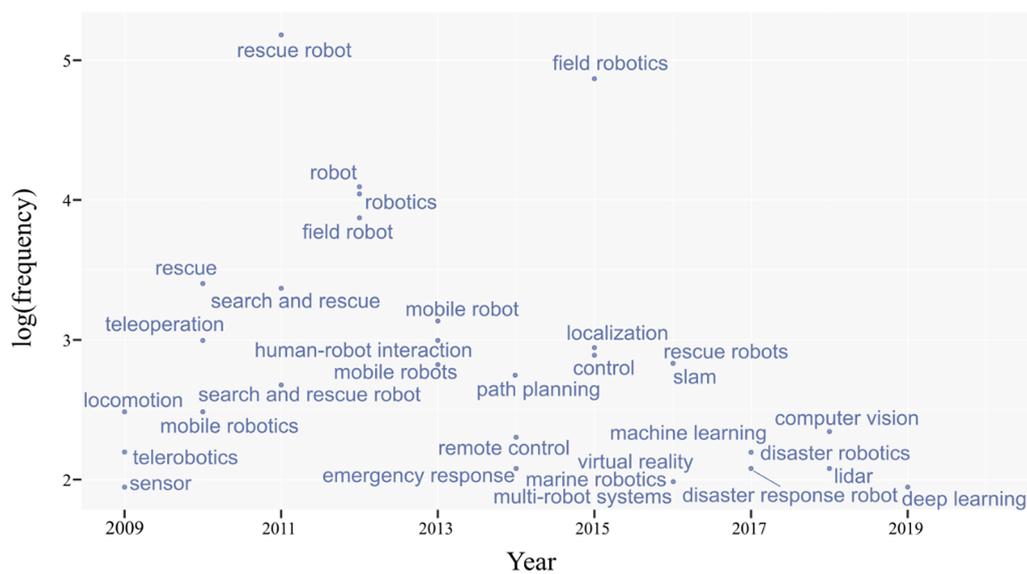

**Fig. 9** A timestamped keyword analysis in RR for the last decade. In this analysis, the top 3 frequent keywords are selected which have been mentioned at least 7 times in the literature during a year

*Fundamental challenges (2009–2012)*

This phase of research in RR, represented by keywords like tele-robotics, tele-operation, sensor, and locomotion, is focused on addressing more fundamental challenges associated with locomotion mechanisms [64], sensor and perception systems [65], and tele-operation [66].

*Control and robotics challenges (2012–2016)*

In the timestamped keyword analysis, the research priorities of RR evolved from fundamental challenges to control and robotics challenges which indicates the priority shift in the field. In this phase, path planning [67], localization, control systems [68], human–robot interaction (HRI) [69], and simultaneous localization and mapping (SLAM) [70] became the research trends of RR.

*AI and machine learning challenges (2016–2020)*

Addressing AI and machine learning challenges can be referred to as the third research trend in RR begetting the second priority shift in the field. This phase is represented by computer science-related keywords like computer vision [71], machine learning [72], visual reality [73], and multi-robot systems [74], which indicate the very same research topics. It is noteworthy that some of these research topics might have been studied even in the early years of RR, but the timestamped analysis is concerned with the global research trends rather than outlier works. For instance, machine learning is a broad and inclusive research topic and has been utilized in RR way earlier than the 2016–2020 period. However, incorporating machine learning techniques in the development of response robots did not attract researchers' attention globally due to the practical and theoretical limitations of the primary generations of response robots.

**Mapping scientific collaboration**

Studying the scientific collaboration mapping reveals the way that prominent authors, leading publications, references, and frequent keywords are related, as Fig. 10 depicts the RR scientific collaboration mapping excluding references. According to the results, *Advanced Robotics* has published a high volume of articles of the prominent authors in RR. As mentioned in Table 5, this journal also ranks first among the top 15 journals in the field of RR which indicates that it is the authors' first preference in general. Among the authors listed in the figure, Tadokoro is the most loyal author to *Advanced Robotics* which emphasizes a stronger correlation between the author's research area and the journal's scope. On the contrary, Brik has published articles in various journals, including *Advanced Robotics*, indicating a broader research area. Moreover, the *Journal of Robotics and Mechatronics* publishes a high volume of articles and ranks 3rd among the top 15 leading journals according to Table 5, though they are associated with a few prominent authors listed in Fig. 10. This figure also maps the authors to the frequently used keywords in publications within the field of RR. The keywords mentioned in this figure that are mutually correlated with the listed authors are inclusive parent keywords from which the specific research areas of authors cannot be discussed. This fact, however, explicitly reiterates the finding that researchers and scholars from a wide range of disciplines have contributed to RR and they have been addressing various challenges within the field; thus, they have introduced numerous technical keywords to the field, and yet the mutual keywords remain the generic ones.

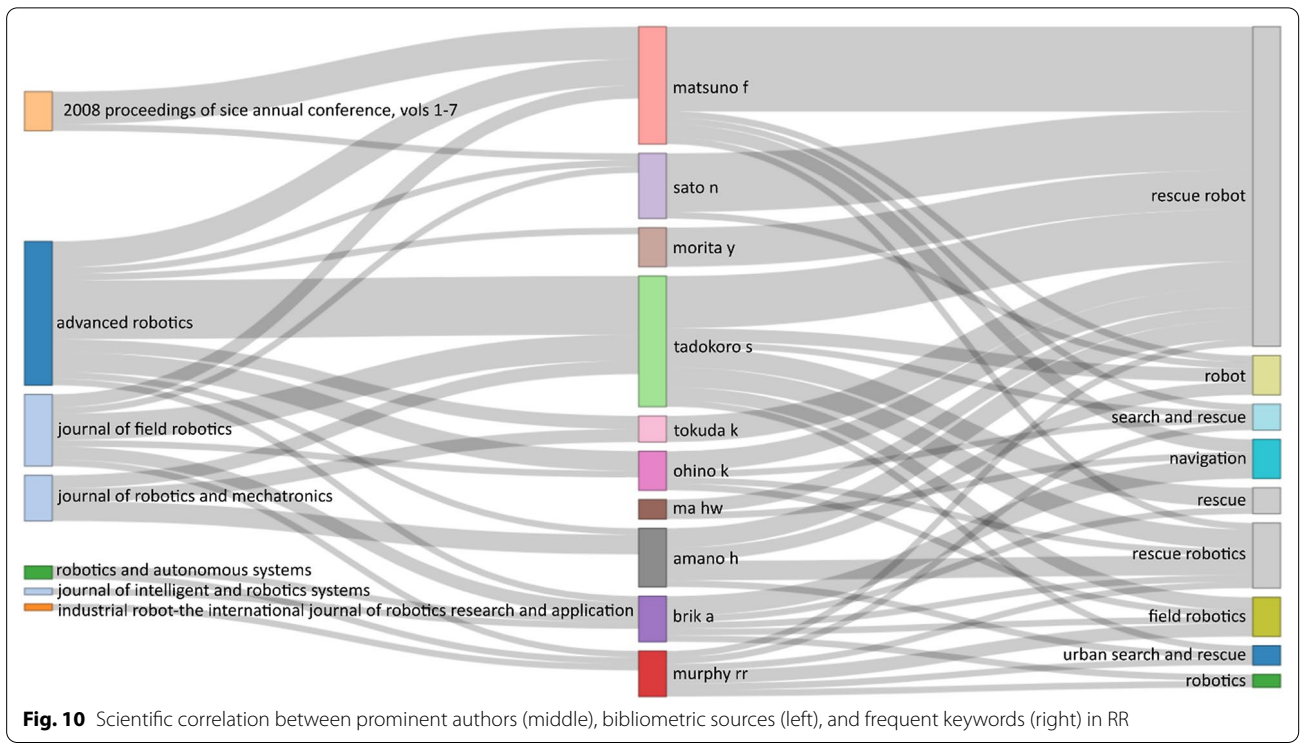

**Fig. 10** Scientific correlation between prominent authors (middle), bibliometric sources (left), and frequent keywords (right) in RR

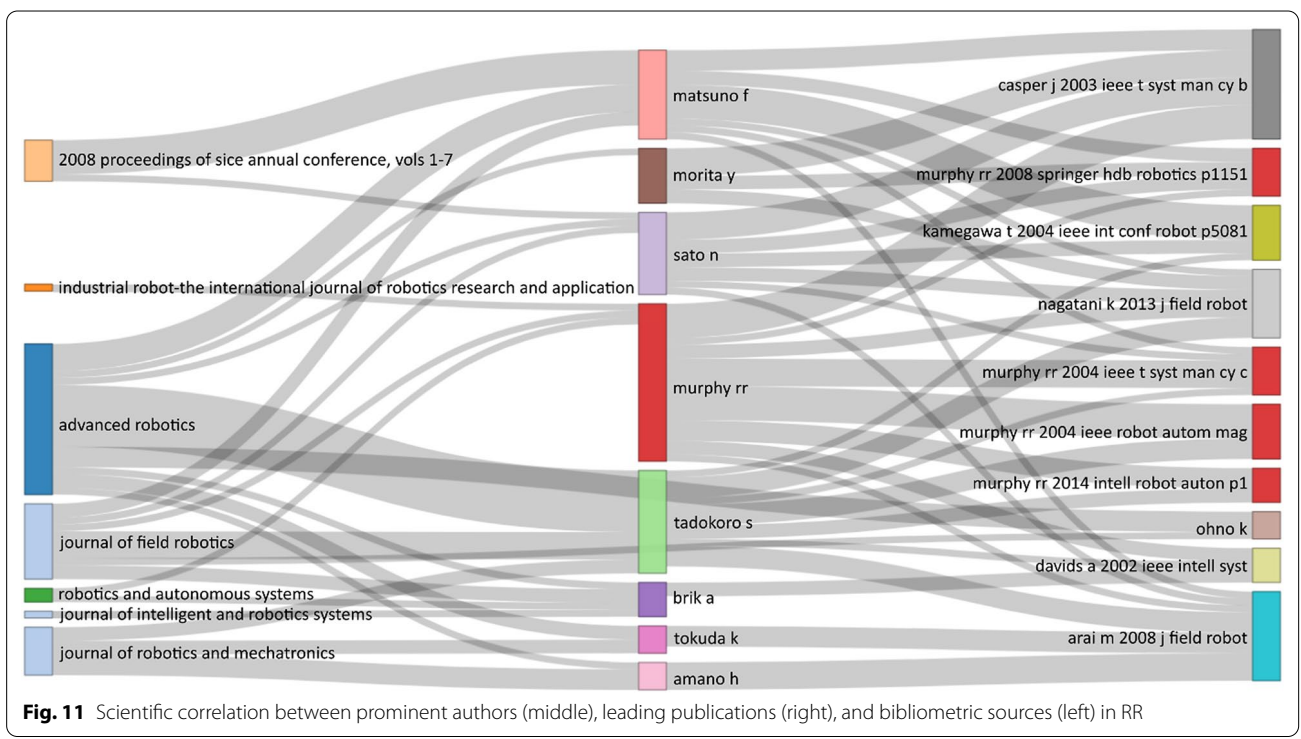

**Fig. 11** Scientific correlation between prominent authors (middle), leading publications (right), and bibliometric sources (left) in RR

The correlation study is reconducted by replacing the keyword column with the reference column in Fig. 10 to investigate the scientific collaboration from another perspective, as shown in Fig. 11. The results of this study demonstrate that most of the prominent authors have utilized various mutual references listed in the figure, though a few of them strictly have used only one reference. For instance, Tokuda and Amano have referred only to [75] in their articles. Moreover, Murphy has authored four articles [11, 42, 76, 77] among 10 references listed in Fig. 11. This scholar is the only prominent author (recognized in "Eminent authors" section) who is also listed as the notable references in Fig. 11. Casper's work on HRI has been most frequently referred to by prominent authors [78]. Although the topic is mainly HRI that is only one of the many sub-fields of RR, it has received tremendous attention from various disciplines because of its application in the world trade center disaster in 2001.

## Discussion

Due to the multidisciplinary nature of RR, researchers and developers of the field incorporated scientific and technological advancements from multiple engineering disciplines into their works. That being the case, several complementary research areas associated with certain leading authors and scholars have been inevitably promoted in RR. This fact explains why leading publications in RR are not necessarily produced by leading authors in the field, discussed in "Leading publications" section. Smith's work on "*The New College Vision and Laser Data Set*," ranked fourth among leading publications in RR in Table 4, exemplifies a complementary research topic/work that has been widely utilized in the RR research community. The multidisciplinary character of RR also explains the diversity between the contributing authors in terms of research areas. This diversity has been manifestly reflected in the clustering displayed in Fig. 3, where the research community is split into nine clusters based on the research areas such as mechanical design and AI/machine learning. From a broader perspective, the multidisciplinary nature of RR led to a wide diversity in the research specialty of the contributing authors in RR.

The diversity of contributing authors led to decent general attention to RR, and yet only a small number of them are consistently and mainly focused on RR. In other words, a noticeable percentage of contributing authors consider RR as an application of their research, while they do not actively contribute to the state-of-the-art advancements in the field. This is mainly because RR is a scientifically attractive and practically justifiable application for a wide variety of research topics in robotics and AI. Is it necessarily an effective contribution to RR? One may underestimate such contributions to RR with no significant scientific implications. Nonetheless, investigating a research topic at the application level appends more practical perspectives to the literature which particularly benefits research and developments of response robots categorized under field robotics. On the whole, authors and scholars who are mainly focused on RR play the most imperative role in expanding the scientific and technological boundaries of the field, though RR benefit from contributing leading authors from complementary research areas.

The core research community of RR, authors who are mainly focused on RR, maintains some properties requiring further studies. Isolated contribution is one noticeable behavior of some leading authors within the community. The isolated contribution does not necessarily imply that the contributing authors overlook the work of the flew scholars. Under this assumption, the lack of overlap in the authors' research area can explain the isolated contribution. Another finding of the core research community is that Murphy can be named the most impactful scholar in RR, as "Eminent authors", "Leading publications" and "Mapping scientific collaboration" sections, additively recognize Murphy's influence and significant contributions. Murphy is recognized as one of the top leading authors with the most impactful contributions whose publications turn out to be the most important references cited by other leading authors. Murphy's impacts are significant since this scholar has been contributing to RR in various research areas including AI and robotics, HRI, and RR-related test and deployment. Aside from Murphy, a considerable number of leading authors are from institutions and agencies in Japan. Generally speaking, authors and scholars from Japan actively contribute to robotics in its various sub-fields including RR. Besides, natural and human-made disasters in Japan called attention to employing state-of-art technologies to lessen the fatalities and better manage such crises. Given these points, RR-focused contributing authors have made significant collective breakthroughs with their various research backgrounds, contribution behaviors, and motivations, which resulted in the notable RR literature during the past decades.

In reference to publication sources, the bibliometric analysis indicates that RR research year publications have been reduced in conferences in the past decade. Despite this reduction, *SSRR* holds the highest total number of publications. Among the journals, on the other hand, *Advanced Robotics* ranked first among the top 15 and has been popular among RR scholars. Additionally, *Advanced Robotics* and the *Journal of Field Robotics* (ranked fifth in top robotics journals) have demonstrated consistent production rates throughout the past decades. It is noteworthy that the *International Journal of Robotics Research* ranked fifth and first between top RR and robotics publishing journals, respectively; it also holds a considerable share (~20%) of total publications in the field of robotics. This shows that RR research owns a particular place between top robotics journals. Overall, the productivity of the top 15 journals has increased since 2005.

The dedicated RR research center at Waseda University manifests the first rank of this institution and its researchers (43 total publications) in the top 15 affiliation list. The publications of authors from Tohoku University place this institution in second place with nearly 7% fewer publications while they received more citations. The contribution of researchers from the University of Sydney and Carnegie Mellon University seems more impactful since almost 80% of their RR research documents were cited. Remarkably, only 29 institutions have more than 10 RR publications, which indicates that RR is only popular among some organizations. Overall, Japanese institutions have demonstrated a dominant presence among the top institutions, which represents Japan's offering to RR and also our assessment about the motivation of Japanese authors.

Regarding the locations, the RR literature has primarily emerged from 12 countries that contributed more than 20 total publications. Following the records obtained from our institution analysis, Japan holds first place in the top 15 countries ranking with 272 RR publications. In fact, Japan can be perceived as the global architecture of RR. Especially, when considering the catastrophic incidents that happened in Japan and urged Japanese researchers to develop robotic solutions to save human lives, whether rescuer or victims in disastrous environments. Additionally, the literature published by Japanese scientists has a low MCP ratio (high SCP), which represents the country as a subdivision in RR. China and the USA are ranked second and third in the top 15 contributing countries, respectively. In contrast, both of these countries demonstrated more international collaborations than Japan with multiple country publications of 24.

Assuredly, the top three countries possess robust funding sources that actively support RR research. According to our analysis of robotics literature, the top funding agencies that support research in robotics are from these countries. American funding agencies (NSF, DARPA, ONR, and ARL) hold the majority in the top 10 list, where NSF-funded publications were cited the most. Based on InCites report, JSPS from Japan has funded the highest number of publications in robotics. On the other hand, the literature funded by the Australian Research Council (ARC) has received the most impactful citations according to WOS data.

The evolution of RR's research priorities has been discussed via bibliometric analysis of frequent keywords in "Analysis of keywords and technical content" section. The research priorities in each phase represent the bottlenecks in advancing RR-related technologies and each priority shift marks momentous progress in addressing the global challenges. The time order of the research priorities reveals fundamental dependencies among different scientific aspects of RR. To put it another way, advancements in some research areas are highly dependent on progress made in other areas, which implies that the time order of RR's research priorities is scientifically sound considering the theoretical and technological dependencies. Considering the current research priorities of RR and the most recent advances in the complementary research areas, the future research priorities and trends of RR can be viewed as:

### Reinforcement learning (RL)

Most of the response operations take place in unknown and dynamic environments which makes any hard-coding and pre-planning infeasible [79, 80]. When no prior experience is available to a decision-making agent (in this case, response robots), RL offers a broad range of model-based and model-free algorithms to improve the agent's performance based on trial and error [58]. Since recent scientific progress in deep RL, hierarchical RL, and generalized RL has not been adequately incorporated into RR yet, addressing learning and planning challenges in RR through RL techniques can potentially be one of the future research priorities in the field.

### Generalized planning

One major challenge in RR is to cope with complicated situations in which all principal capabilities of a response robot are required at a time to accomplish the assigned task. In these situations, a robot's planning would be susceptible to converge to sub-optimal solutions considering the dynamic, partially observable, and stochastic nature of disastrous environments. From an AI and planning point of view, one applicable solution for these situations is: planning for a more straightforward setting with abstracted states and actions first and then generalizing the solution to more complicated scenarios. This paradigm of planning [81] can be a research priority in RR when response robots are being widely deployed in various missions and customizing the response robots' decision-making structure for each mission is not an option due to limitations of time and expert human resources.

**Smart assistive rescue equipment**
Reviewing the RR's global research trends during the past decades indicates that the contributing researchers have been mainly exploring mobile robotic solutions to address formidable challenges in disastrous environments. Considering the contemporary breakthroughs in extended reality, exoskeleton, embedded graphical processing unit (GPU), and machine learning and AI [82, 83], smart assistive equipment can potentially be RR's research priority in the future.

**Heterogeneous multi-agent system**
Employing a group of response robots with a different set of capabilities introduces complex planning and control challenges. Response operations mainly require a group of robots with diverse capabilities and physical properties to successfully be executed. Although the problem of multi-agent planning and control in RR have been theoretically studied based on simulation environments, heterogeneous multi-agent systems [84] are expected to receive tremendous attention from the RR community as response robots are gradually being employed more frequently in response operations.

## Conclusion
RR is a multidisciplinary field that attracts scientists with diverse research backgrounds since it can be a potential application for every robotic solution. This work explores various aspects of this field from its early days through a bibliometric analysis. Since the field is widely spread among countries, institutions, and research communities, we centralized the study on the eminent authors and their publications, and studied the network between them. Moreover, we classified the associated institutions and countries that are pioneers in navigating the research in RR. The RR research within the field of robotics was also studied from the standpoint of the publishing sources and their association with funding agencies.

Our analysis indicates that Japan's researchers and institutions are leading RR globally. Their remarkable contribution to RR originates from the country's history with the Great East Japan Earthquake that intrinsically motivated the research. Furthermore, our bibliometric analysis suggests that the global research focus has begun to grow in new directions within the scientific community since 2016. The recent approaches go beyond the system design problems and investigate more complex operations with intelligent response robots in proximity to humans or even in a group of robots. Understandably, the wide range of AI applications has been leveraged in this transformation. Motivated by the results of this work, we discussed the future trends of RR research that concurrently growing within various domains of AI.

### Abbreviations
RR: Response robotics; WOS: Web of science; JCR: Journal Citation Reports; SLAM: Simultaneous localization and mapping; UGV: Unmanned ground vehicle; UAV: Unmanned aerial vehicle; ROV: Remotely operated underwater vehicle; HRI: Human–robot interaction; USAR: Urban search and rescue; AI: Artificial Intelligence; SICE: Instrument and Control Engineers of Japan; SSRR: IEEE International Workshop on Safety Security and Rescue Robotics; IROS: IEEE/RSJ International Conference on Intelligent Robots and Systems; ROBIO: IEEE International Conference on Robotics and Biomimetics; MCP: Multiple country citation; SCP: Single country citation; TP: Total publications; TC: Total citations; ACPP: Average citation per publication; TCPY: Total citation per year; RL: Reinforcement learning; TC: Times cited; TD: Total WOS documents; IF: Impact factor; GPU: Graphical processing unit; %DC: Percent of WOS documents cited; CI: Citation impact; CMU: Carnegie Mellon University; USYD: University of Sydney; NSF: National Science Foundation; JSPS: Japan Society for the Promotion of Science; MEXT: Ministry of Education, Culture, Sports, Science and Technology, Japan; ARC: Australian Research Council; NSERC: Natural Sciences and Engineering Research Council of Canada; DARPA: Defense Advanced Research Projects Agency; ONR: Office of Naval Research; ARL: US Army Research Lab; NSFC: National Natural Science Foundation of China; NIST: National Institute of Standard and Technology.

### Acknowledgements
Not applicable.

### Authors' contributions
MD and SH conducted research, analyzed the data, developed the key studies, and wrote the manuscript. All authors read and approved the final manuscript.

### Funding
This research received no specific Grant from any funding agency in the public, commercial, or not-for-profit sectors.

###  Availability of data and materials
The data that support the findings of this study are retrieved from the Web of Science, Clarivate Analytics (WOS) database.

### Declarations

### Ethics approval and consent to participate
Not applicable.

### Consent for publication
Not applicable.

### Competing interests
The authors declare that they have no competing interests.

### Author details
[1]Was with Department of Computer Science, Lamar University, Beaumont, USA. [2]Was with Department of Mechanical Engineering, Bucknell University, Lewisburg, USA.

**Publisher's Note**

Springer Nature remains neutral with regard to jurisdictional claims in published maps and institutional affiliations.